\documentclass[conference]{IEEEtran}
\IEEEoverridecommandlockouts
\usepackage{cite}
\usepackage{amsmath,amssymb,amsfonts}
\usepackage{algorithmic}
\usepackage{graphicx}
\usepackage{textcomp}

\usepackage{amssymb}
\usepackage{pifont}

\usepackage{xcolor}
\def\BibTeX{{\rm B\kern-.05em{\sc i\kern-.025em b}\kern-.08em
    T\kern-.1667em\lower.7ex\hbox{E}\kern-.125emX}}
\begin{document}

\title{Security Methods in Internet of vehicles}

\author{\IEEEauthorblockN{1\textsuperscript{st} Mahmood A. Al-Shareeda}
\IEEEauthorblockA{\textit{National Advanced IPv6 Centre} \\
\textit{Universiti Sains Malaysia}\\
Penang, Malaysia}
\and
\IEEEauthorblockN{2\textsuperscript{nd} Selvakumar Manickam}
\IEEEauthorblockA{\textit{National Advanced IPv6 Centre} \\
\textit{Universiti Sains Malaysia}\\
Penang, Malaysia }
}

\maketitle

\begin{abstract}
The emerging wireless communication technology known as vehicle ad hoc networks (VANETs) has the potential to both lower the risk of auto accidents caused by drivers and offer a wide range of entertainment amenities. The messages broadcast by a vehicle may be impacted by security threats due to the open-access nature of VANETs. Because of this, VANET is susceptible to security and privacy problems. In order to go beyond the obstacle, we investigate and review some existing researches to secure communication in VANET. Additionally, we provide overview, components in VANET in details.

\end{abstract}


\section{Introduction}

The improvement of road transportation is the goal of VANET. According to a UK government report on traffic accidents in 2015, there were 1,732 fatalities and 22,137 injuries. It is challenging to control traffic on crowded city streets, and doing so frequently results in unpleasant events like traffic jams, accidents, fuel waste, wasted time, etc. As a result, it is now essential to safely and effectively manage traffic on highways in these cities \cite{cui2020edge,cui2018efficient,cui2017spacf}. The fast developing wireless communication technologies enable effective management of the traffic generated by thousands of cars, enabling intelligent transportation systems (ITS). Equipment known as on-board units (OBUs) can be put on vehicles thanks to ITS applications that take the form of vehicular ad-hoc networks (VANETs)\cite{zhang2019pa,al2021security,zhong2019privacy}.

\begin{figure}[h]
	\includegraphics[width=.45\textwidth]{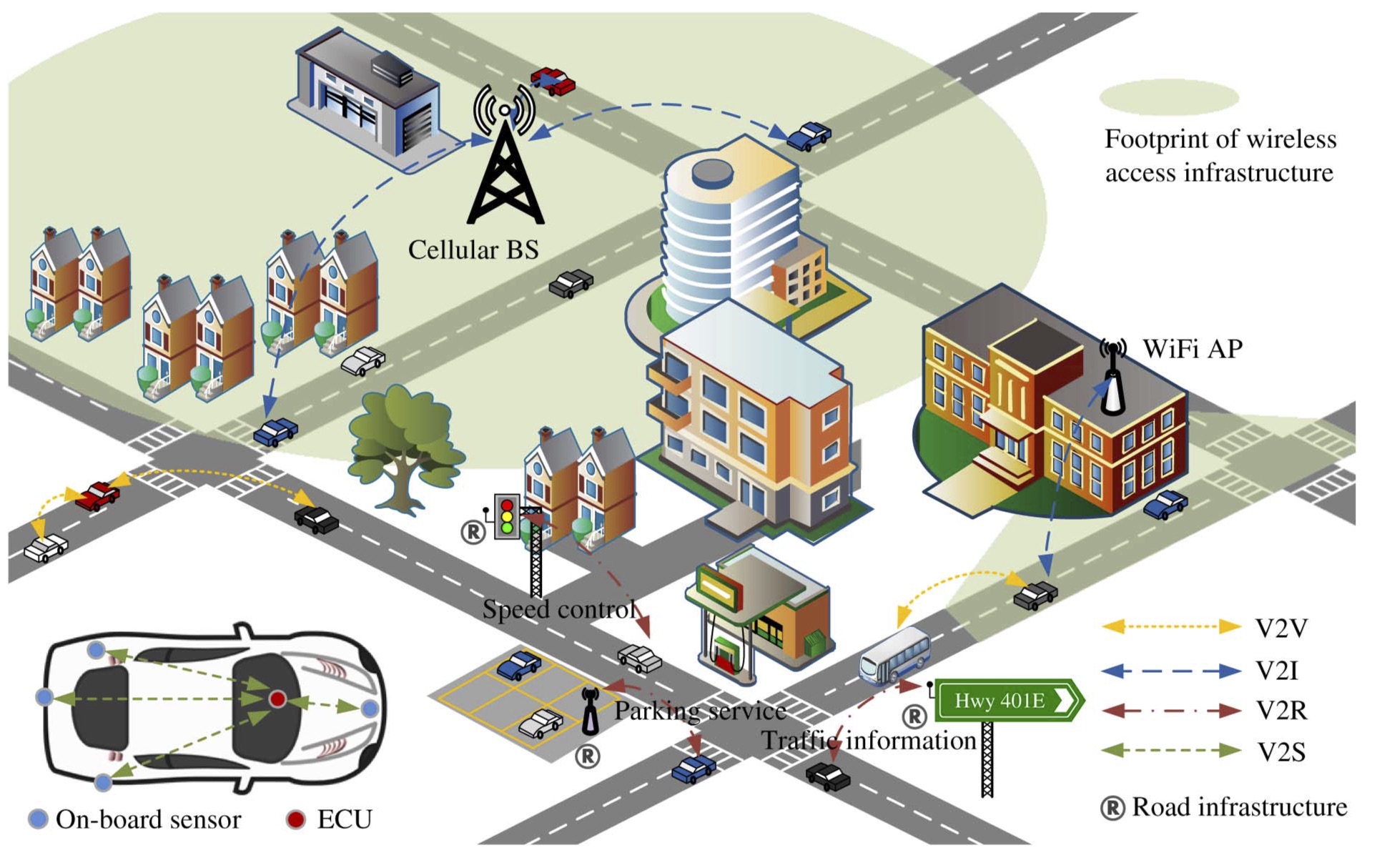}
	\caption{Environment of Safety Road \cite{viloria2020vanet}.} \label{ddd}
\end{figure}

Through a dedicated short range communications (DSRC) system, the OBU in each vehicle transmits safety messages to the OBUs in nearby cars and infrastructure (such as the road-side unit (RSU) and the trusted authority (TA)) within a range of 300 metres, as shown in Figure \ref{ddd}.
The US Department of Transportation's groundbreaking ITS effort for VANET standardisation is called the DSRC. Both vehicle-to-vehicle (V2V) and vehicle-to-infrastructure (V2I) communication are supported by VANETs. For the purpose of preventing collisions, reducing traffic congestion, and enhancing the driving environment, communicative vehicles (V2V) communicate information about their speed, location, heading, and traffic jams with one another.

Due to their crucial significance in the field of smart transportation by supporting vehicle-to-vehicle and vehicle-to-infrastructure communication, vehicular ad hoc networks (VANETs) have become more widespread in recent years \cite{al2020survey,al2020security}.

\section{Vehicular Ad hoc Network (VANET)}
\subsection{Overview}
A vehicular ad hoc network (VANET) contains a group of automobiles, moving or stationary, that are connected by a wireless network. VANETs were developed with the intention of giving drivers comfort and safety in moving vehicles. This viewpoint is evolving since VANETs are now seen as the foundation for smart transportation systems that enable autonomous vehicles and any activity that needs Internet connection in the context of a smart city environment. Additionally, VANETs allow onboard computers in stationary vehicles, such as those in airport parking lots, to function as mobile computing cloud resources with the least amount of Internet infrastructure support.

\subsection{VANET Components}
\begin{figure}[h]
	\includegraphics[width=.45\textwidth]{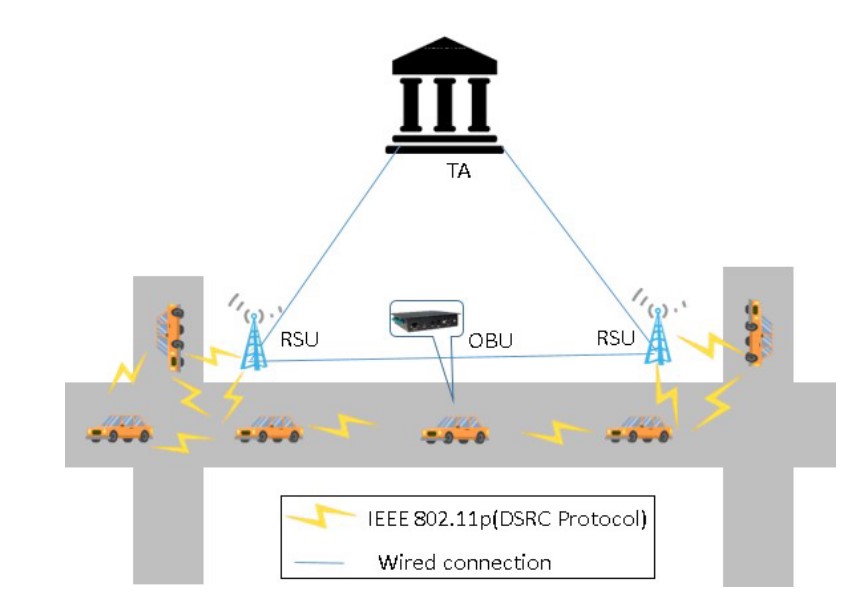}
	\caption{Typical VANETs.} \label{phasesvanet123}
\end{figure}

As shown in Figure \ref{phasesvanet123}, A typical VANETs consists of the following three primary parts:
\begin{itemize}

\item OBU: Every vehicle has a wireless OBU that allows for the processing, transmission, and reception of beacons using the DSRC protocol.

     \item RSU: It is situated by the side of the road and is tasked with managing each OBU within its coverage area using the DSRC protocol. The RSU communicates with OBUs and TA via wireless networks and a secure wired network, respectively. A Tamper-Proof Device (TPD) is a component of each RSU that is used to store private information and perform system-wide cryptography operations. Therefore, the opponent is unable to disclose it.
      
    \item TA: Is in charge of registering vehicles, RSUs, and generating system parameters in VANETs. It is also in charge of handling the revocation procedure.
       
\end{itemize}

\section{Exciting Research}

In order to solve the issues listed above in VANET, \cite{al2020efficientideinit}  suggested a solution in this study called efficient conditional privacy preservation with mutual authentication. This system is predicated on the partition and distribution of geographical regions into a number of domains, each of which is responsible for storing the Certificate Revocation List (CRL) in all Road-side Units (RSUs) situated inside the domain. The vehicle should authenticate with the TA during the mutual authentication phase. The vehicle is permitted to send a message to the other components in the VANET after receiving a pool of pseudo-identities and the matching secret keys from RSU.

\cite{al2020vppcs}  offered a privacy-preserving communication scheme (VPPCS) based on the VANET that satisfies the needs for both content and contextual privacy. It makes use of identity-based encryption and elliptic curve cryptography (ECC). To confirm and verify the suggested system, we performed a thorough security analysis using the Burrows-Abadi-Needham (BAN) logic, the random oracle model, the security of proof, and security attributes.

In order to discover an optimal or almost optimal solution, \cite{al2018towards} provided the well-known genetic algorithm based on RSU location in this study. We offer the fundamental simulation environment needed for this job. To produce automotive mobility, simulate road traffic, execute vehicular network simulations on Omnet++ using the veins model framework, simulate realistic networks using Omnet++, and build the algorithm to analyse the results, we need to use OSM to get real map data, GatcomSUMO, Omnet++, and Matlab.

In this research, \cite{al2020efficient} offered a conditional identity-based privacy-preserving authentication technique that provides batch verification for the concurrent message verification by each node. Additionally, the TPD regularly and often updates its database of vehicle information to thwart side-channel attacks.

To find the most suitable or nearly optimal solution, \cite{al2019realistic} employed the well-known genetic algorithm that is mostly based on the RSU region. We provide the basic simulation environment for this work using OpenStreetMap (OSM) to download real map data, Group for Architecture and Computer Technology (Gatcom) to generate car mobility, Software Update Monitor (SUMO) to simulate street traffic, Veins model framework for simulating walking vehicular networks, OMNET++ to simulate real networks, and Matlab to create the algorithm to analyse the results.

\cite{al2020review} investigated and examined the impact of the replay attack, one of the most frequent security assaults in VANETs. This paper covers the findings from the research and analysis, as well as an examination of the replay attack defence mechanisms already in use in VANETs.

\cite{hamdi2020techniques} talked about what a traffic monitoring centre is and how it works. Additionally, we will describe the idea of early event detection and highlight the drawbacks and advantages of each method.

\cite{al2020lswbvm} suggested an Elliptic Curve Cryptography-based Lightweight Security Without Batch Verification Method (LSWBVM) technique (ECC). During mutual authentication, the proposed LSWBVM technique employs the XOR operation and universal hash function. The proposed system, however, leverages an effective single verification rather than batch verification in a high traffic density location to check a large number of messages. The LSWBVM accomplishes the security objectives of mutual authentication between the nodes, as shown by BAN logic. The security study shows that the LSWBVM satisfies the security requirements, including safe non-forgery under the random oracle model in an adaptively generated message assault, identity privacy preservation, tractability, and revocability.

The fundamental backdrop of VANET from an architectural perspective and communication kinds are covered in  \cite{al2020modreview}. The overview of the VANET modification attack is then given. This report also presents a comprehensive evaluation of the current VANET modification attack preventive measures. This review study shows that a more effective preventive strategy is still required to deal with the modification assault in VANET.

Two theoretical models that display and predict the QoS in terms of performance are presented in \cite{hamdi2020performance}. In terms of throughput, the star architecture performs 15\% better than the grid layout. While the average end-to-end latency record for star network topology and grid network topology, respectively, is 17.25 ms and 76.25 ms. Grid and star topologies both see a reduction in jitter as the data rate increases from 1 to 11 Mbps; however, in all testing, the average jitter was lower in star topologies than in grids.

According to the literature, \cite{al2020newreview} discovered many Man-In-The-Middle (MITM) attacks with a range of behaviours, including message tampering, message stalling, and message dropping. The fundamental backdrop of VANET from an architectural perspective and communication kinds are covered in this paper. The outline of the MITM attack in the VANET is then given. This study also presents a comprehensive evaluation of the current VANET MITM attack protection measures. This review study shows that a more effective preventive strategy is still required to deal with the MITM attack in VANET.

Any link within VANETs must be protected from cyberattacks, which is a need \cite{alazzawi2020id}. This is because data transmission in any context with open access could result in a variety of network assaults. To address many problems relating to security and privacy in VANETs, the identity-based privacy-preserving authentication technique (ID-PPA) was presented in this study. Numerous ID-based security protocols for VANETs have recently been introduced in \cite{alazzawi2020id}.

On order to secure communication in VANETs, \cite{al2020ne} suggested the New and Efficient Conditional Privacy-Preserving Authentication (NE-CPPA) scheme. Elliptic Curve Cryptography (ECC) is the algorithm that NE-CPPA utilises to meet security and privacy criteria. The suggested scheme's security analysis is presented for several attack scenarios.

In \cite{al2021se}, a Secure and Efficient Conditional Privacy-Preserving Authentication (SE-CPPA) technique is developed for thwarting impersonation attacks and improving performance efficiency. The bilinear pair cryptography used for message signing and verification forms the foundation of the SE-CPPA proposal. The proposed SE-CPPA method can achieve security goals in terms of formal and informal analysis through security analysis and comparison. The genuine identification of the vehicle contained in the tamper-proof device (TPD) is often updated and has a limited time of validity in order to fend off impersonation attacks.

On order to protect communication in VANETs, \cite{al2022secure} suggested a secure pseudonym-based conditional privacy-persevering authentication system. The proposed method for message signing and verification used safe hash cryptography and elliptic curve cryptography (ECC). During the broadcasting process, a vehicle employs a sizable number of pseudo-IDs it has received from the Trusted Authority (TA) and the appropriate signature key to sign messages. As a result, the proposed approach mandates that every vehicle examine each broadcasting message that it receives. Additionally, the proposed approach allows the TA to stop misbehaving vehicles from broadcasting signed messages continually, protecting against insider attacks.

 \cite{al2022password} suggested a Chinese Remainder Theorem (CRT)-based password-guessing attack-aware authentication method that would protect inter-vehicle communication on 5G-enabled vehicular networks. In the suggested method, trusted authorities (TAs) generate and broadcast fresh group keys to the CRT-assisted vehicles. Furthermore, the suggested approach simply needs believable TPDs because the system's master key does not have to be preloaded. The suggested method prevents password guessing attacks and ensures the highest level of security for all 5G-enabled vehicle networks.

\cite{al2021towards}  examined the security flaws in the current schemes. In order to secure and boost the effectiveness of VANETs communications, it also suggests improvements to the identity-based conditional privacy-preserving authentication mechanism. The suggested approach has been shown to be secure using the random oracle model in addition to meeting the security and privacy requirements.

The design and development of any new VANETs approaches may use in \cite{al2021comprehensive} as a guide and a reference. Additionally, this paper may aid academics and developers in choosing the key VANET characteristics for their objectives in a single document.

An effective conditional privacy-preserving authentication (E-CPPA) system based on elliptic curve encryption is presented in \cite{al2020proposed}. In the VANETs system, E-CPPA increases the effectiveness of message signing and verification for both vehicle to infrastructure and vehicle to vehicle communication. The E-CPPA scheme's goal is to satisfy demands about the security and privacy of VANETs. Finally, this study discusses the anticipated outcome for the E-CPPA scheme and subsequent research in addition to discussing the critical review of comparable works in VANETs.

In the context of 5G-enabled vehicle networks, \cite{al2022chebyshev} suggested a Chebyshev polynomial-based method for fending off side-channel attacks. Our study was able to accomplish the Chebyshev polynomial's crucial chaotic and semi-group features. System startup, enrollment, signature, verification, and pseudonym renewal are the five stages of our job. Additionally, our work regularly and constantly updates the vehicle's information in the TPD to fend off side-channel attacks.

In order to ensure communication security in 5G-enabled automotive networks, a chaotic map-based conditional privacy-preserving authentication (CM-CPPA) technique is suggested in \cite{s22135026}. The proposed CM-CPPA method uses a chaotic map-based hash function and a Chebyshev polynomial mapping operation to sign and validate communications. Additionally, the suggested CM-CPPA scheme's security analysis results utilising the AVISPA simulator are good and secure against common threats.

\section{Conclusion}\label{sec:con}

VANET can improve traffic efficiency and safety by fostering cooperative communication between cars, roadside infrastructure, and traffic management centres. To guarantee safe service delivery in VANET, message authentication is essential. This work has examined and analysed some recent studies on communication security in VANETs.

	\bibliographystyle{unsrt}
	\bibliography{Main}

\begin{thebibliography}{10}

\bibitem{cui2020edge}
Jie Cui, Lu~Wei, Hong Zhong, Jing Zhang, Yan Xu, and Lu~Liu.
\newblock Edge computing in vanets-an efficient and privacy-preserving
  cooperative downloading scheme.
\newblock {\em IEEE Journal on Selected Areas in Communications},
  38(6):1191--1204, 2020.

\bibitem{cui2018efficient}
Jie Cui, Jing Zhang, Hong Zhong, Runhua Shi, and Yan Xu.
\newblock An efficient certificateless aggregate signature without pairings for
  vehicular ad hoc networks.
\newblock {\em Information Sciences}, 451:1--15, 2018.

\bibitem{cui2017spacf}
Jie Cui, Jing Zhang, Hong Zhong, and Yan Xu.
\newblock Spacf: A secure privacy-preserving authentication scheme for vanet
  with cuckoo filter.
\newblock {\em IEEE transactions on vehicular technology}, 66(11):10283--10295,
  2017.

\bibitem{zhang2019pa}
Jing Zhang, Jie Cui, Hong Zhong, Zhili Chen, and Lu~Liu.
\newblock Pa-crt: Chinese remainder theorem based conditional
  privacy-preserving authentication scheme in vehicular ad-hoc networks.
\newblock {\em IEEE Transactions on Dependable and Secure Computing},
  18(2):722--735, 2019.

\bibitem{al2021security}
Mahmood~A Al-Shareeda, Mohammed Anbar, Selvakumar Manickam, Ayman Khalil, and
  Iznan~Husainy Hasbullah.
\newblock Security and privacy schemes in vehicular ad-hoc network with
  identity-based cryptography approach: A survey.
\newblock {\em IEEE Access}, 9:121522--121531, 2021.

\bibitem{zhong2019privacy}
Hong Zhong, Shunshun Han, Jie Cui, Jing Zhang, and Yan Xu.
\newblock Privacy-preserving authentication scheme with full aggregation in
  vanet.
\newblock {\em Information Sciences}, 476:211--221, 2019.

\bibitem{viloria2020vanet}
Amelec Viloria, Omar Bonerge~P{\'\i}neda Lezama, and Noel Varela.
\newblock Vanet heterogeneous networks with wireless technology variation
  according to the capacity of users in urban areas.
\newblock {\em Journal of Intelligent \& Fuzzy Systems}, 39(6):8325--8332,
  2020.

\bibitem{al2020survey}
Mahmood~A Al-Shareeda, Mohammed Anbar, Iznan~Husainy Hasbullah, and Selvakumar
  Manickam.
\newblock Survey of authentication and privacy schemes in vehicular ad hoc
  networks.
\newblock {\em IEEE Sensors Journal}, 21(2):2422--2433, 2020.

\bibitem{al2020security}
MAASM Mahmood~A Al-shareeda, Mohammed Anbar, Murtadha~A Alazzawi, Selvakumar
  Manickam, and Iznan~H Hasbullah.
\newblock Security schemes based conditional privacy-preserving in vehicular ad
  hoc networks.
\newblock {\em Indonesian Journal of Electrical Engineering and Computer
  Science}, 21(1), 2020.

\bibitem{al2020efficientideinit}
Mahmood~A Al-shareeda, Mohammed Anbar, Selvakumar Manickam, and Iznan~H
  Hasbullah.
\newblock An efficient identity-based conditional privacy-preserving
  authentication scheme for secure communication in a vehicular ad hoc network.
\newblock {\em Symmetry}, 12(10):1687, 2020.

\bibitem{al2020vppcs}
Mahmood~A Al-Shareeda, Mohammed Anbar, Selvakumar Manickam, and Ali~A Yassin.
\newblock Vppcs: Vanet-based privacy-preserving communication scheme.
\newblock {\em IEEE Access}, 8:150914--150928, 2020.

\bibitem{al2018towards}
Mahmoud Al~Shareeda, Ayman Khalil, and Walid Fahs.
\newblock Towards the optimization of road side unit placement using genetic
  algorithm.
\newblock In {\em 2018 International Arab Conference on Information Technology
  (ACIT)}, pages 1--5. IEEE, 2018.

\bibitem{al2020efficient}
Mahmood~A Al-Shareeda, Mohammed Anbar, Iznan~Husainy Hasbullah, Selvakumar
  Manickam, and Sabri~M Hanshi.
\newblock Efficient conditional privacy preservation with mutual authentication
  in vehicular ad hoc networks.
\newblock {\em IEEE Access}, 8:144957--144968, 2020.

\bibitem{al2019realistic}
Mahmoud Al~Shareeda, Ayman Khalil, and Walid Fahs.
\newblock Realistic heterogeneous genetic-based rsu placement solution for v2i
  networks.
\newblock {\em Int. Arab J. Inf. Technol.}, 16(3A):540--547, 2019.

\bibitem{al2020review}
Mahmood~A Al-shareeda, Mohammed Anbar, Iznan~H Hasbullah, Selvakumar Manickam,
  Nibras Abdullah, and Mustafa~Maad Hamdi.
\newblock Review of prevention schemes for replay attack in vehicular ad hoc
  networks (vanets).
\newblock In {\em 2020 IEEE 3rd International Conference on Information
  Communication and Signal Processing (ICICSP)}, pages 394--398. IEEE, 2020.

\bibitem{hamdi2020techniques}
Mustafa~Maad Hamdi, Lukman Audah, Sami~Abduljabbar Rashid, and Mahmood
  Al~Shareeda.
\newblock Techniques of early incident detection and traffic monitoring centre
  in vanets: A review.
\newblock {\em J. Commun.}, 15(12):896--904, 2020.

\bibitem{al2020lswbvm}
Mahmood~A Al-Shareeda, Mohammed Anbar, Murtadha~A Alazzawi, Selvakumar
  Manickam, and Ahmed~Shakir Al-Hiti.
\newblock Lswbvm: A lightweight security without using batch verification
  method scheme for a vehicle ad hoc network.
\newblock {\em IEEE Access}, 8:170507--170518, 2020.

\bibitem{al2020modreview}
Mahmood~A Al-shareeda, Mohammed Anbar, Selvakumar Manickam, and Iznan~H
  Hasbullah.
\newblock Review of prevention schemes for modification attack in vehicular ad
  hoc networks.
\newblock {\em International Journal of Engineering and Management Research},
  10, 2020.

\bibitem{hamdi2020performance}
Mustafa~Maad Hamdi, Ahmed~Shamil Mustafa, Hussain~Falih Mahd, Mohammed~Salah
  Abood, Chanakya Kumar, and Mahmood~A Al-shareeda.
\newblock Performance analysis of qos in manet based on ieee 802.11 b.
\newblock In {\em 2020 IEEE international conference for innovation in
  technology (INOCON)}, pages 1--5. IEEE, 2020.

\bibitem{al2020newreview}
Mahmood~A Al-shareeda, Mohammed Anbar, Selvakumar Manickam, and Iznan~H
  Hasbullah.
\newblock Review of prevention schemes for man-in-the-middle (mitm) attack in
  vehicular ad hoc networks.
\newblock {\em International Journal of Engineering and Management Research},
  10, 2020.

\bibitem{alazzawi2020id}
Murtadha~A Alazzawi, Hasanain~AH Al-behadili, Mohsin~N Srayyih~Almalki,
  Aqeel~Luaibi Challoob, and Mahmood~A Al-shareeda.
\newblock Id-ppa: robust identity-based privacy-preserving authentication
  scheme for a vehicular ad-hoc network.
\newblock In {\em International Conference on Advances in Cyber Security},
  pages 80--94. Springer, 2020.

\bibitem{al2020ne}
Mahmood~A Al-shareeda, Mohammed Anbar, Selvakumar Manickam, Iznan~H Hasbullah,
  Nibras Abdullah, Mustafa~Maad Hamdi, and Ahmed~Shakir Al-Hiti.
\newblock Ne-cppa: A new and efficient conditional privacy-preserving
  authentication scheme for vehicular ad hoc networks (vanets).
\newblock {\em Appl. Math}, 14(6):1--10, 2020.

\bibitem{al2021se}
Mahmood~A Al-Shareeda, Mohammed Anbar, Selvakumar Manickam, and Iznan~H
  Hasbullah.
\newblock Se-cppa: A secure and efficient conditional privacy-preserving
  authentication scheme in vehicular ad-hoc networks.
\newblock {\em Sensors}, 21(24):8206, 2021.

\bibitem{al2022secure}
Mahmood~A Al-Shareeda, Mohammed Anbar, Selvakumar Manickam, and Iznan~H
  Hasbullah.
\newblock A secure pseudonym-based conditional privacy-preservation
  authentication scheme in vehicular ad hoc networks.
\newblock {\em Sensors}, 22(5):1696, 2022.

\bibitem{al2022password}
Mahmood~A Al-Shareeda, Mohammed Anbar, Selvakumar Manickam, and Iznan~H
  Hasbullah.
\newblock Password-guessing attack-aware authentication scheme based on chinese
  remainder theorem for 5g-enabled vehicular networks.
\newblock {\em Applied Sciences}, 12(3):1383, 2022.

\bibitem{al2021towards}
Mahmood~A Al-Shareeda, Mohammed Anbar, Selvakumar Manickam, and Iznan~H
  Hasbullah.
\newblock Towards identity-based conditional privacy-preserving authentication
  scheme for vehicular ad hoc networks.
\newblock {\em IEEE Access}, 2021.

\bibitem{al2021comprehensive}
Mahmood~A Al-shareeda, Murtadha~A Alazzawi, Mohammed Anbar, Selvakumar
  Manickam, and Ahmed~K Al-Ani.
\newblock A comprehensive survey on vehicular ad hoc networks (vanets).
\newblock In {\em 2021 International Conference on Advanced Computer
  Applications (ACA)}, pages 156--160. IEEE, 2021.

\bibitem{al2020proposed}
Mahmood~A Al-shareeda, Mohammed Anbar, Selvakumar Manickam, Iznan~H Hasbullah,
  Ayman Khalil, Murtadha~A Alazzawi, and Ahmed~Shakir Al-Hiti.
\newblock Proposed efficient conditional privacy-preserving authentication
  scheme for v2v and v2i communications based on elliptic curve cryptography in
  vehicular ad hoc networks.
\newblock In {\em International Conference on Advances in Cyber Security},
  pages 588--603. Springer, 2020.

\bibitem{al2022chebyshev}
Mahmood~A Al-Shareeda, Selvakumar Manickam, Badiea~Abdulkarem Mohammed,
  Zeyad~Ghaleb Al-Mekhlafi, Amjad Qtaish, Abdullah~J Alzahrani, Gharbi
  Alshammari, Amer~A Sallam, and Khalil Almekhlafi.
\newblock Chebyshev polynomial-based scheme for resisting side-channel attacks
  in 5g-enabled vehicular networks.
\newblock {\em Applied Sciences}, 12(12):5939, 2022.

\bibitem{s22135026}
Mahmood~A. Al-Shareeda, Selvakumar Manickam, Badiea~Abdulkarem Mohammed,
  Zeyad~Ghaleb Al-Mekhlafi, Amjad Qtaish, Abdullah~J. Alzahrani, Gharbi
  Alshammari, Amer~A. Sallam, and Khalil Almekhlafi.
\newblock Cm-cppa: Chaotic map-based conditional privacy-preserving
  authentication scheme in 5g-enabled vehicular networks.
\newblock {\em Sensors}, 22(13), 2022.

\end{thebibliography}
\end{document}